\documentclass[showpacs,twocolumn,showemail]{revtex4}
\usepackage{bbm}
\usepackage{mathrsfs}
\usepackage{epsfig,psfrag}
\usepackage{amsmath,amsfonts,amssymb}
\usepackage[usenames]{color}
\usepackage[dvips, bookmarks, colorlinks=true, plainpages = false, citecolor = blue, linkcolor = blue, urlcolor = blue, filecolor = blue]{hyperref}
\def\be{\begin{equation}}
\def\ee{\end{equation}}
\def\bea{\begin{eqnarray}}
\def\eea{\end{eqnarray}}

\begin{document}

\preprint{draft}

\title{Thermal Behavior of Spin Clusters and Interfaces in two-dimensional Ising Model on Square Lattice }
\author{  Abbas Ali Saberi }
\email{a$_$saberi@ipm.ir}

\address {School of Physics, Institute for Research in Fundamental
Sciences (IPM), P.O.Box 19395-5531, Tehran, Iran}
\date{\today}

\begin{abstract}
Extensive Monte Carlo study of two-dimensional Ising model is done
to investigate the statistical behavior of spin clusters and
interfaces as a function of temperature, $T$. We use a
\emph{tie-breaking} rule to define interfaces of spin clusters on
square lattice with strip geometry and show that such definition is
consistent with conformal invariant properties of interfaces at
critical temperature, $T_c$. The \emph{effective} fractal dimensions
of spin clusters and interfaces ($d_c$ and $d_I$, respectively) are
obtained as a function of temperature. We find that the effective
fractal dimension of the spin clusters behaves almost linearly with
temperature in three different regimes. It is also found that the
effective fractal dimension of the interfaces undergoes a sharp
crossover around $T_c$, between values $1$ and $1.75$ at low and
high temperatures, respectively. We also check the finite-size
scaling hypothesis for the percolation probability and the average
mass of the largest spin-cluster in a good agreement with the
theoretical predictions.
\end{abstract}


\maketitle

Two dimensional (2D) Ising model as a solvable prescription model in
hand, and its extension to \emph{q}-state Potts model \cite{Potts}
have been the subject of intense research interest for decades. Many
of their thermodynamical parameters and behaviors can be
characterized in terms of some fractal geometrical objects, e.g.,
spin clusters and domain walls. Most of studies have been focused to
describe the behavior of these models at critical temperature $T_c$,
at which they exhibit a continuous phase transition (for
$\emph{q}\leq4$), and less attention is made to investigate
off-critical characterization of such systems at temperatures far
from $T_c$. At $T=T_c$, conformal field theory (CFT) plays an
important robust role to describe the universal critical properties
in two dimensions. Besides CFT, theory of stochastic Loewner
evolution (SLE) invented by Schramm \cite{schramm} provides a
geometrical understanding of criticality which states that the
statistics of well defined domain walls (or curves, e.g., spin
cluster boundaries in 2D Ising model) in upper half plane
$\mathbb{H}$ is governed by one-dimensional Brownian motion (to
review SLE, see \cite{SLE}). Therefore it is expected that for
example in 2D Ising model, the geometrical exponents such as the
fractal dimension of a spin cluster and its boundary as well would
be related to the thermodynamical exponents \cite{janke}. The study
of the fractal structure and the scaling properties of the various
geometrical features of the Ising model has been subject of huge
scientific literature (see for example \cite{NC, BH, Coniglio,
Duplantier, SV} and references therein). It is also well known that
most two-dimensional critical models renormalize onto a Gaussian
free field theory (Coulomb gas) \cite{LPK}. Many exact critical
exponents have been computed by using the Coulomb gas technique
\cite{BN}. These include various geometrical exponents of
two-dimensional Ising
model \cite{BKN}, and general \emph{q}-state Potts model \cite{BN2, DBN}. \\
The geometrical objects reflect directly the status of the system in
question under changing the controller parameters. Temperature can
play the role of such controller parameter in 2D Ising model.

Investigation of the dependence of geometrical exponents in 2D Ising
model, equivalent to $q=2$ states Potts model, is the main subject
of the present paper. To be consistent with the postulates of SLE at
$T=T_c$, we consider the model on a strip of size $L_x\times L_y$,
where $L_x$ is taken to be much larger than $L_y=L$, i.e., $L_x=8L$.
We simulate the spin configurations of 2D Ising model on square
lattice using Wolf's Monte Carlo algorithm \cite{wolf}, based on
single cluster update. Before going into the further details, let us
address an ambiguity that arises when one intends to define an Ising
interface on a square lattice, and then introduce a rule which seems
to produce well-defined interfaces on square lattice.
\begin{figure}[b]\begin{center}
\includegraphics[scale=0.6]{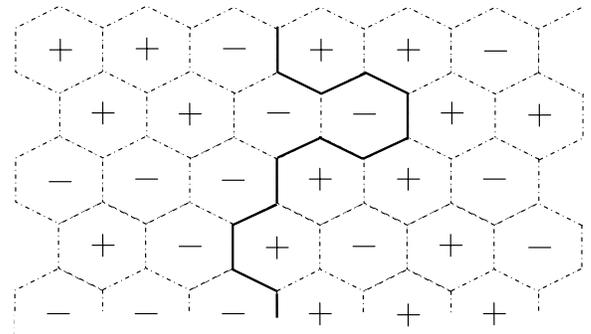}
\narrowtext\caption{\label{Fig1}An Ising interface defined on
hexagonal lattice corresponding to a spin configuration on
triangular lattice with a fixed boundary condition at the real line
in $\mathbb{H}$, as explained in the text.}\end{center}
\end{figure}
The importance of such definition backs to its relevance to the SLE
interfaces at criticality. As will be discussed later, it is
believed that the critical Ising interfaces can be defined by the
theory of SLE in the scaling limit \cite{smirnov, cardy}. Thus we
need to have a unique procedure to define operationally the hulls of
the Ising spin clusters without any self intersection and ambiguity.
However we will use our following procedure to define well defined
interfaces in Ising spin model, one can simply extend it for any
two-dimensional model defined on square lattice, e.g., for
interfaces of general \emph{q}-state Potts model or contour lines of
random growth surfaces \cite{SRR} etc, with appropriate
substitutions of spins up and down.

Consider an Ising model on a triangular lattice in upper half plane
on which each spin lies at the center of a hexagon having six
nearest neighbors and the spin boundaries (defining the interface)
lie on the edges of the honeycomb lattice (see Fig. \ref{Fig1}). To
impose an interface (which separates the spins of opposite
magnetization), growing from the origin on the real line to
infinity, a fixed boundary condition can be considered in which all
spins in the right and left sides of the origin are up ('$+$') and
down ('$-$'), respectively. The Gibbs distribution induces a measure
on these interfaces.\\To define an interface, a walker moves on the
edges of the hexagonal lattice starting from origin at the bottom.
At each step the walker moves according to the following rule: turns
left or right according to the value of the spin in front of it
('$+$' or '$-$', respectively). The resulting interface is a unique
interface which never crosses itself and never gets trapped. Such a
interface, at $T=T_c$, is believed to be described by SLE in the
continuum limit \cite{smirnov, cardy}.

This procedure to define the interface should be modified for spin
configuration on square lattice. This is because that there are some
choices for the square lattice, at places with four alternating
spins. We first introduce a \emph{tie-breaking} rule which the
walker regards at each step and then we show that this definition is
consistent with the predictions of SLE for such interfaces at
$T=T_c$.

\begin{figure}[t]\begin{center}
\includegraphics[scale=0.7]{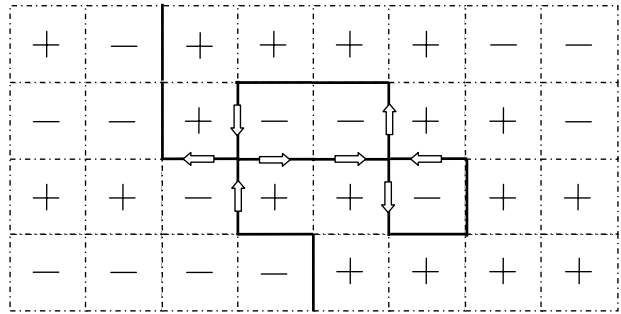}
\narrowtext\caption{\label{Fig2}An Ising interface defined on square
lattice, dual of the original square lattice including a spin
configuration, with a fixed boundary condition at the real line in
$\mathbb{H}$. The interface is generated applying the turn-right
\emph{tie-breaking} rule. The same procedure can be used to define
such interface for down spins ('$-$') according to turn-left
\emph{tie-breaking} rule.}\end{center}
\end{figure}

Consider a spin configuration on a strip of square lattice in
$\mathbb{H}$, with the same boundary conditions as above. A walker
moves along the edges of the dual lattice (the lattice shown by the
dotted-dashed lines in the Fig. \ref{Fig2}), starting from the
origin. According to the boundary conditions at the first step of
the walk that the spins '$+$' lie at the \emph{right} of the walker,
this direction is chosen to be the preferable direction. After
arriving to each site on the dual lattice, there are three
possibilities for the walker: it can cross one of the three nearest
bonds of the original lattice. At the first step of selection, it
chooses the bonds containing two different spins that crossing each
of which leaves the spin '$+$' at the \emph{right} and spin '$-$' at
the \emph{left} of the walker. The  directions \emph{right} and
\emph{left} are defined locally according to the orientation of the
walker. After the first selection, if there are yet two
possibilities to cross, the walker chooses the bond which accords
with the turn-right \emph{tie breaking} rule: it turns towards the
bond which is in its right hand side with respect to its last
direction at the last walk; if there is not any selected bond at its
right, it prefers to move straightly and if there is not also any,
it turns to its left. The procedure is repeated iteratively until
the walker touches the upper boundary. The resulting interface is
again an interface which touches itself yet never crosses itself and
never gets trapped. The same procedure can be used to define another
interface with \emph{left}-preferable direction as turn-left
\emph{tie-breaking} rule.\\ It would be worth to mention that the
procedure introduced here yields not just a unique cluster boundary
without any ambiguity on the square lattice, but one can check that,
any other definition for the interface leads to an incorrect
boundary of the cluster (for example at vertices with more than one
possibility, just these introduced options lead to the 'true'
boundary of the considered cluster and any other option, for example
choosing randomly the directions left or right, may enter the
boundary of a spin which does not belong to the cluster. Note that a
spin cluster is defined as a set of nearest neighbor connected sites
of like sign.).

\begin{figure}[b]\begin{center}
\includegraphics[scale=0.45]{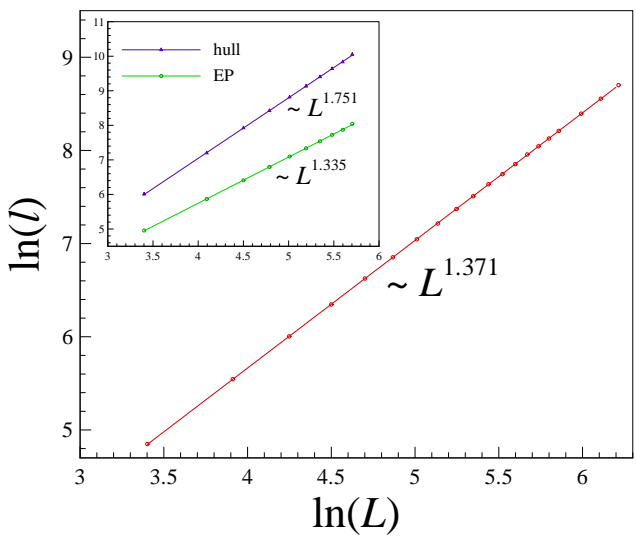}
\narrowtext\caption{\label{Fig3} (Color online) Log-log plot of the
average length of a spanning interface $l$, generated using the
\emph{tie-breaking} rule introduced in the text, versus the wide of
the strip $L$, at critical points. Main: for Ising model. Inset: for
the hull (the upper graph) and its external perimeter (EP$-$the
lower graph) of critical site percolation. The values of the best
fit to the data are represented aside each one, with an error of
$\sim0.005$. }\end{center}
\end{figure}

Let us now show that the resulting interface is compatible with the
properties which comes from their conformal invariant nature at
$T=T_c$.\\The Wolf's Monte Carlo algorithm is used to simulate the
spin configurations at $T=T_c$, on the strip of square lattice and
of aspect ratio $8$, and boundary conditions as discussed above. For
each size $L$, about $4L^2$ Monte Carlo sweeps are used for
equilibration. An ensemble of $2\times10^4$ independent samples is
collected for each sample size $L$, where each of which was taken
after $10L$ Monte Carlo steps.\\Each spin cluster has been
identified as a set of connected sites of the same spin using
Hoshen-Kopelman algorithm. We just take the samples including a
vertical spanning cluster in the \emph{y}-direction. Then an
ensemble of corresponding spanning interfaces was obtained using
mentioned turn-right (left) \emph{tie-breaking} rule.\\The fractal
dimension of the interfaces at this critical temperature,
$d_I(T_c)$, is obtained using the standard finite size scaling. The
length of an interface $\emph{l}$ scales with the sample size as
$\emph{l}\sim L^{d_I(T_c)}$. The fractal dimension of conformally
invariant curves is provided by SLE \cite{SLE} generally as $d_I =
1+\kappa/8$, where diffusivity $\kappa$ classifies different
universality classes, and for Ising spin-cluster boundaries it is
conjectured to be $\kappa=3$ and thus $d_I(T_c)=\frac{11}{8}=1.375$.
As shown in Fig. \ref{Fig3}, the best fit to our data collected for
sizes $30\leq L\leq 500$ yields the fractal dimension
$d_I(T_c)=1.371\pm0.005$.

\begin{figure}[t]\begin{center}
\includegraphics[scale=0.45]{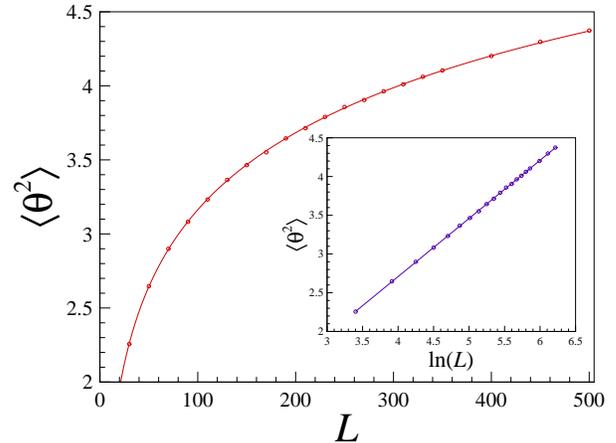}
\narrowtext\caption{\label{Fig4}(Color online) Variance of the
winding angle for spanning interfaces generated using the
\emph{tie-breaking} rule introduced in the text. The solid line is
set according to the Eq. (\ref{winding}), with $a=-0.29$ and
$\kappa=3$. In the inset, the variance in semilogarithmic
coordinates. }\end{center}
\end{figure}

\begin{figure*}[t]\begin{center}
\includegraphics[scale=0.52]{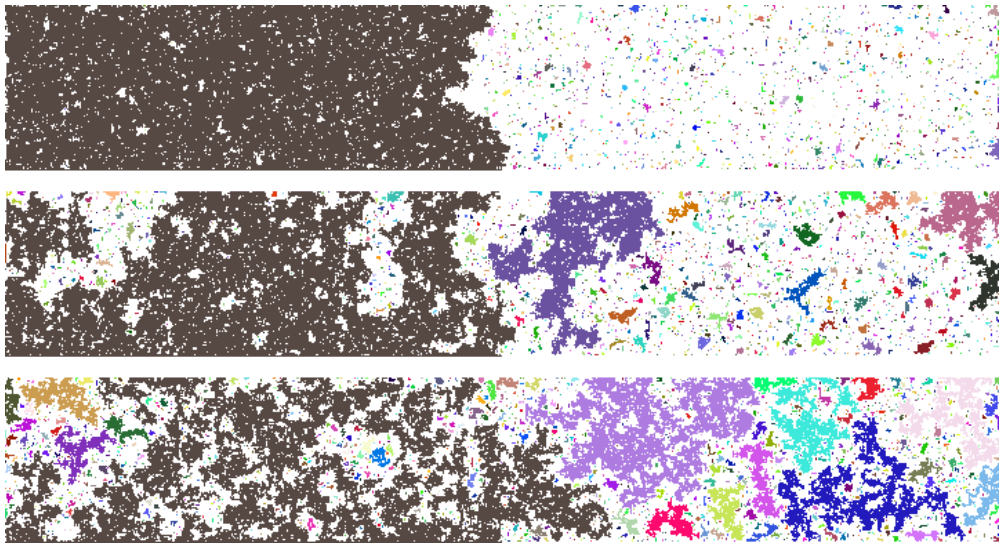}
\narrowtext\caption{\label{Fig5}(Color online) Spin clusters of 2D
Ising model on strip of square lattice with size of $L=120$, and
aspect ratio $6$, at different temperatures from top to bottom:
$T-T_c = -0.2, 0$ and, $0.2$. The boundary conditions (bc) used for
simulation are fixed for the lower boundary, antiperiodic at sides
and free bc for upper one. The spin-down clusters are shown white.
The bc imposes an interface at the boundary of the spanning cluster
(dark colored) starting from the origin (using the turn-left
\emph{tie-breaking} rule in these figures) and ending at the upper
boundary. As temperature increases the interface gets more space
filling.}\end{center}
\end{figure*}
\begin{figure}[h]\begin{center}
\includegraphics[scale=0.35]{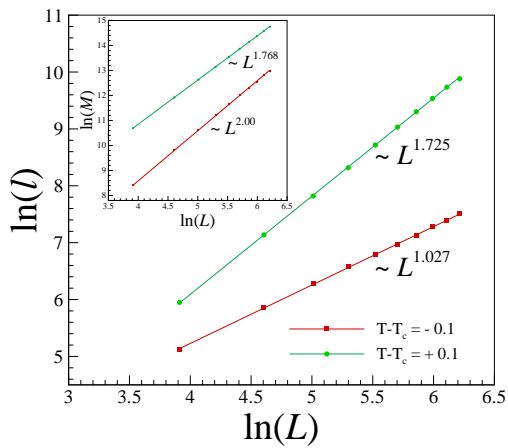}
\narrowtext\caption{\label{Fig6}(Color online) Main: Log-log plot of
the average length of a spanning interface $l$, versus the wide of
the strip $L$, at two different temperatures $T-T_c=-0.1$ and
$+0.1$. These graphs show the scaling property and the fractal
behavior of the interfaces far from criticality at length scales
less than the correlation length. Inset: Log-log plot of the average
mass of a spanning cluster $M$, versus the wide of the strip $L$, at
$T-T_c=-0.1$ and $+0.1$. The graph for $T-T_c=+0.1$ is shifted
upwards by $2$.}\end{center}
\end{figure}

Another prediction of the theory of SLE for such critical interfaces
is the winding angle statistics \cite{schramm}. We define the
winding angle $\theta$ as defined by Wieland and Wilson
\cite{wilson}. For each interface we attribute an arbitrary winding
angle to the first edge (that we take zero). Then the winding angle
for the next edge is defined as the sum of the winding angle of the
present edge and the turning angle to the new edge measured in
radians. It is shown that \cite{wilson, duplantier} the variance in
the winding grows with the sample size like \be\label{winding}
\langle\theta^2\rangle = a+ \frac{\kappa}{4}\ln L,\ee where
$\kappa=8[d_I(T_c)-1]$, and $a$ is a constant whose value is
irrelevant. So the exact value of $\kappa$ for critical interfaces
of 2D Ising model should be $\kappa=3$.
\\The figure \ref{Fig4} indicates that our result for $\kappa$ is in
a good agreement with the predicted value. We find that
$\kappa=3.012\pm0.005$.\\We have also tested other conformal
invariant properties of the interfaces such as Schramm's formula for
the left passage probability of the interfaces, consistent with the
theory (the results are not shown here).

To investigate another concern about the systems with more
complicated interfaces, we did such experiments for the critical
site percolation \cite{lee}. The fractal dimension of the hull and
its external perimeter are obtained as $d^{H}_I=1.751\pm0.002$, and
$d^{EP}_I=1.335\pm0.002$, respectively (see Fig. \ref{Fig3}) in a
good agreement with the duality relation predicted from the
conformal invariant property \cite{duplantier0} \be\label{duality}
(d^H_I-1)(d^{EP}_I-1) = \frac{1}{4}.\ee

In the rest of the paper, let us consider the statistical
geometrical response of the Ising model to the temperature. We show
experimentally that how the statistics of the spin clusters and
their boundaries behave as a function of temperature. We try to
measure the corresponding fractal dimensions at length scales
smaller than the correlation length $\xi$, using the standard finite
size scaling as done at critical temperature above.\\ Features of
the spin clusters at three different temperatures are shown in Fig.
\ref{Fig5}. These represent what we expect to happen: at zero
temperature, because of the used boundary conditions, the ground
state of the spin configuration splits the system into two segments,
one with spins up and the other with spins down which are separated
with a straight interface. Increasing in the temperature induces a
fractal random feature on spin clusters and interfaces. The
interfaces are some non-intersecting curves (in $\mathbb{H}$) which
can be described, in the continuum limit, by a dynamical process
called Loewner evolution \cite{Loewner} with a suitable continuous
driving function $\zeta_t$ as \be \label{Loewner} \frac{\partial
g_t(z)}{\partial t}=\frac{2}{g_t(z)-\zeta_t}, \ee where, if we
consider the hull $K_t$, the union of the curve and the set of
points which can not be reached from infinity without intersecting
the curve, then $g_t(z)$ is an analytic function which maps
$\mathbb{H}\setminus K_t$ into the $\mathbb{H}$ itself.\\At zero
temperature the driving function $\zeta_t$, is an specific constant,
at $T=T_c$ it should be proportional to a standard Brownian motion
$B_t$ as $\zeta_t=\sqrt{\kappa}B_t$ with $\kappa=3$, and it may be
complicated random function at other different temperatures.\\ At
high-temperature limit, each spin gets the directions up or down
with probability $p=1/2$ and so, it is conjectured to correspond to
the critical site percolation on triangular lattice (on which the
percolation threshold is exactly at $p_c=1/2$), and it is expected
that the driving function converges to a Brownian motion with
diffusivity of $\kappa=6$. For the case of square lattice, since the
percolation threshold in two dimensions is at $p_c\sim 0.59$, so at
high-temperature limit where $p=\frac{1}{2}< p_c$, the system will
be below the threshold and the crossover to the critical site
percolation will not be seen any more.

\begin{figure}[b]\begin{center}
\includegraphics[scale=0.35]{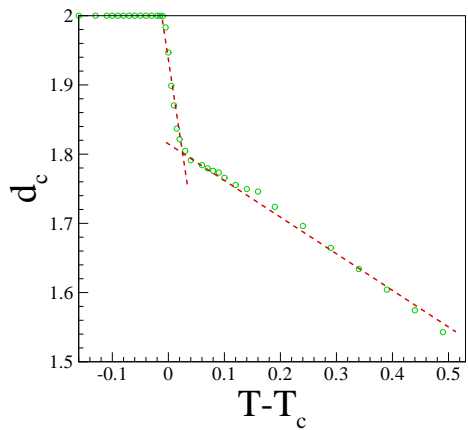}
\narrowtext\caption{\label{Fig7}(Color online) Effective fractal
dimension of spin clusters as a function of temperature. It changes
almost linearly in three different regimes: low temperature with
dimension of $2$, rapid decreasing around $T_c$ and a crossover to a
different linear behavior far from $T_c$. The slope of the
dashed-lines differs by one order of magnitude. Each point is
obtained using finite size scaling for $10$ different sizes in the
range of $50\leq L\leq 500$. The error is less than the symbol
size.}\end{center}
\end{figure}

\begin{figure}[t]\begin{center}
\includegraphics[scale=0.37]{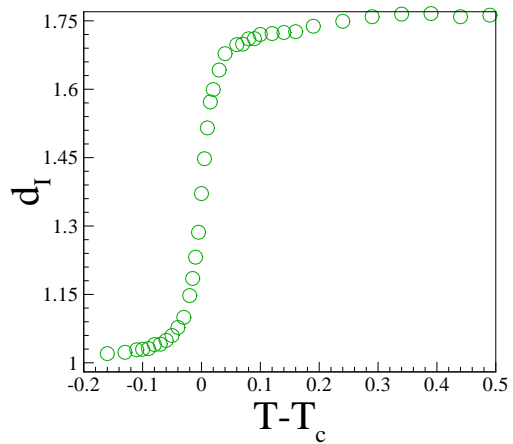}
\narrowtext\caption{\label{Fig8}(Color online) Effective fractal
dimension of spin cluster boundaries as a function of temperature.
The error is less than the symbol size.}\end{center}
\end{figure}

Before looking at the temperature dependence of the fractal
dimension of the spin-clusters, let us discuss more about their
scaling properties from the point of view of theoretical
expectations.\\ The Ising model is expected to be scale-invariant
(on scales much larger than lattice spacing \emph{a}) only at
renormalisation group fixed points, i.e., $T=T_c$ and $T=\infty$ (on
triangular lattice). At those points one expects well-defined
power-law behavior for clusters and their hulls on all scales
$L\gg\emph{a}$. For $T$ just above $T_c$, where the correlation
length $\xi$ is finite and $\xi\gg\emph{a}$, one expects to see
behavior characteristic of the critical point $T_c$ on scales
$\emph{a}\ll L\ll \xi$, and of the high-temperature fixed point on
scales $L\gg \xi$.\\Thus, according to the theory, there should be
no such thing as '\emph{the fractal dimension at temperature} $T$',
except for $T=T_c$ and $T=\infty$, instead one should see a
crossover between two different values. If one chooses a
sufficiently narrow range of length scales one will see an
\emph{effective} fractal dimension, which will have the appearance
of depending on temperature. However, for the Ising model on square
lattice, since the crossover to the critical percolation at
high-temperatures no longer exists, the behavior of the effective
fractal dimensions is governed by just the behavior at $T=T_c$ for
length scales $\emph{a}\ll L\ll \xi$. In order to determine the
behavior of such effective fractal dimensions as a function of
temperature, we measure them in an almost narrow range of sizes $L$,
which seem to be much smaller than the correlation length and within
the range the scaling properties are held.

Figure \ref{Fig6} shows the procedure we perform to measure the
effective fractal dimension of the spin clusters and interfaces at
different temperatures. The finite size scaling reduces
substantially the statistical errors in estimating the fractal
dimensions. The average is taken over $10^4$ independent samples of
aspect ratio $4$, at each temperature below $T_c$ for each sample
size (only the spanning cluster in each configuration and the
corresponding interface was considered). Since the probability to
have a spanning cluster diminishes when temperature increases (as
will be discussed later), the average is taken over $2\times10^4$
independent samples for $T>T_c$, and the samples were
gathered on strip of aspect ratio $8$.\\
The exact values for the fractal dimensions of spin clusters and
interfaces are known just for at critical temperature $T_c$, as
$d_c(T_c)=\frac{187}{96}=1.9479...$ and
$d_I(T_c)=\frac{11}{8}=1.375$, respectively. Our measurements of
fractal dimensions at $T_c$ which give $d_c(T_c)=1.9469\pm 0.001$
and $d_I(T_c)=1.371\pm0.005$ are in a good agreement with the exact
results. These values were obtained for $30\leq L\leq 500$. The same
measurements for $T\neq T_c$, were done for $10$ different sizes
within $50\leq L\leq 500$ (the examples are shown in Fig. \ref{Fig6}).\\
Fig. \ref{Fig6} shows the scaling properties and the fractal
behavior of the spin clusters and interfaces at $T\neq T_c$, within
the selected range of size.

To quantify the geometrical changes of the spin clusters at
different temperatures, we measure the effective fractal dimensions
of the spin clusters and their perimeters. At each temperature, we
use the scaling relation between the average mass of the spanning
spin-cluster $M$, and the width of the strip $L$, to measure the
fractal dimension of the spin-clusters $-$ i.e., $M\sim L^{d_c}$.

Corresponding fractal dimension of spin clusters as a function of
temperature is shown in Fig. \ref{Fig7}. This suggests three
different regimes, one for low temperatures in which the dimension
of the spin clusters is $2$. The second regime is in the vicinity of
the critical temperature: a linear dependence of the fractal
dimension on temperature with a sharp decreasing which is governed
by criticality. A crossover happens at temperature above critical
region which changes the slope of the linear decrease by about one
order of magnitude at high temperatures.

Such a crossover can be also seen in the behavior of the effective
fractal dimension of the interfaces as a function of temperature. As
shown in Fig. \ref{Fig8}, at low temperatures the effective fractal
dimension of the interfaces is close to $1$ and it increases with
temperature. In the vicinity of the critical temperature it
increases again sharply and then crosses over to the value very
close to $1.75$, which is the fractal dimension of the hull of
critical percolation. The whole behavior looks like a hyperbolic
tangent function.
\begin{figure}[b]\begin{center}
\includegraphics[scale=0.37]{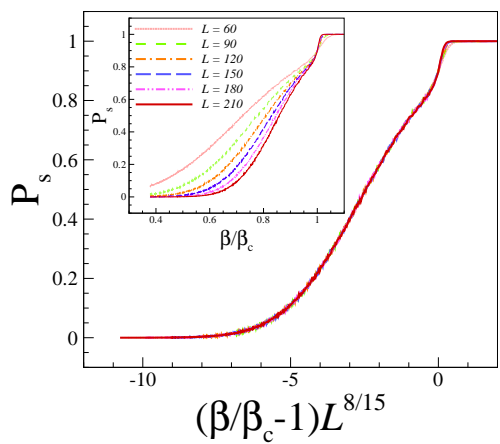}
\narrowtext\caption{\label{Fig9}(Color online) Finite-size scaling
plots of the data for the percolation probability, measured on
square lattices of different size $L^2$. Inset: the percolation
probability as a function of the inverse temperature
$\beta$.}\end{center}
\end{figure}
The other theoretical predictions for the geometrical features
considered in this paper and we are interested in checking them, are
about the percolation observables. The finite-size scaling
hypothesis states that the percolation probability $P_s$ i.e., the
probability to have a spanning cluster at temperature $T$, reaching
from one boundary to the opposite one, behaves like \cite{BH}
\be\label{P_s}P_s=P_s(L/\xi), \ee where the correlation length
behaves like $\xi \sim(T-T_c)^{-\nu}$, with $\nu=15/8$ for the Ising
spin geometric clusters.\\ In order to investigate this hypothesis,
we have done simulations of Ising model on square lattices of
different size $L^2$ with free boundary condition, and the
measurements are taken by averaging over $2\times10^4$ independent
samples at each temperature. As shown in Fig. \ref{Fig9}, curves
$P_s$ measured on lattices of different size all cross at the
critical point (in the figure this observable is shown as a function
of the inverse temperature $\beta$). As can be seen from the figure,
applying the scaling theory Eq. \ref{P_s}, results data collapse
onto a single function, in a good agreement with the theoretical
predictions.

\begin{figure}[h]\begin{center}
\includegraphics[scale=0.35]{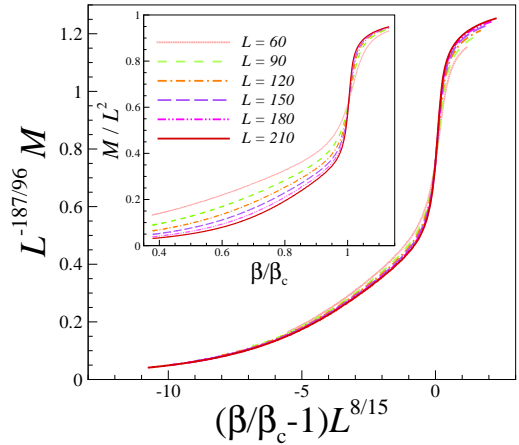}
\narrowtext\caption{\label{Fig10}(Color online) Data collapse for
the average mass of the largest spin-cluster $M$, measured on square
lattices of different size $L^2$. Inset: the strength of the largest
spin-cluster as a function of the inverse temperature
$\beta$.}\end{center}
\end{figure}

The other observable we consider is the scaling behavior of the
average mass of the largest spin-cluster, $M$. According to theory,
this should have the scaling form \be M=L^{d_c(T_c)} F(L/\xi),\ee
where the scaling function $F(x)$ goes to a constant as
$x\rightarrow0$ (at $T=T_c$).
\\ The suitably rescaled mass of the largest spin-cluster as a function of
the reduced inverse temperature is plotted in Fig. \ref{Fig10},
implying the data collapse onto a universal curve.

In conclusion, we studied the geometrical changes of the spin
clusters and interfaces of two-dimensional Ising model on square
lattice in the absence of external magnetic field, as a function of
temperature. We introduced a well-defined \emph{tie-breaking} rule
to generate nonintersecting interfaces on square lattice, which are
shown to be consistent with the predictions of conformal invariance
at the critical point. The results are also checked for critical
site percolation in a good agreement with the analytical predictions.\\
We also investigated the effect of the temperature on the
statistical properties of geometrical objects by measuring the
\emph{effective} fractal dimensions of the spin clusters and
interfaces as a function of temperature. We showed that a crossover
happens which distinguishes between the behavior of these
geometrical objects near the critical temperature and that of at
high temperatures.\\We also applied the finite-size scaling
hypothesis for both the percolation probability and the average mass
of the largest spin-cluster, and we found a data collapse onto a
universal curve, in a good agreement with the theoretical
predictions.

I am indebted to J. Cardy and D.B. Wilson for electronic discussions
and their constructive comments. I also would like to thank H.
Dashti-Naserabadi for his helps on programming.

\end{document}